\documentclass[twocolumn,showpacs,superscriptaddress,preprintnumbers,amsmath,amssymb]{revtex4}
\usepackage{graphics,epsfig,subfigure}
\usepackage{color}

\usepackage[colorlinks=true, linkcolor=red, citecolor=blue]{hyperref}

\begin{document}
\newcommand{\PR}[1]{\ensuremath{\left[#1\right]}} 
\newcommand{\PC}[1]{\ensuremath{\left(#1\right)}} 
\newcommand{\PX}[1]{\ensuremath{\left\lbrace#1\right\rbrace}} 
\newcommand{\BR}[1]{\ensuremath{\left\langle#1\right\vert}} 
\newcommand{\KT}[1]{\ensuremath{\left\vert#1\right\rangle}} 
\newcommand{\MD}[1]{\ensuremath{\left\vert#1\right\vert}} 

\title{Shadow and weak deflection angle of a black hole in nonlocal gravity}
\author{Qi-Ming Fu}
\email{fuqiming@snut.edu.cn}
\affiliation{Department of Physics, College of Sciences, Northeastern University, Shenyang 110819, China}
\affiliation{Institute of Physics, Shaanxi University of Technology, Hanzhong 723000, China}
\author{Shao-Wen Wei}
\email{weishw@lzu.edu.cn}
\affiliation{Institute of Theoretical Physics and Research Center of Gravitation, Lanzhou University, Lanzhou 730000, China}
\affiliation{Lanzhou Center for Theoretical Physics, Key Laboratory of Theoretical Physics of Gansu Province, School of Physical Science and Technology, Lanzhou University, Lanzhou 730000, China}
\author{Li Zhao}
\email{lizhao@lzu.edu.cn}
\affiliation{Institute of Theoretical Physics and Research Center of Gravitation, Lanzhou University, Lanzhou 730000, China}
\affiliation{Lanzhou Center for Theoretical Physics, Key Laboratory of Theoretical Physics of Gansu Province, School of Physical Science and Technology, Lanzhou University, Lanzhou 730000, China}
\author{Yu-Xiao Liu}
\email{liuyx@lzu.edu.cn}
\affiliation{Institute of Theoretical Physics and Research Center of Gravitation, Lanzhou University, Lanzhou 730000, China}
\affiliation{Lanzhou Center for Theoretical Physics, Key Laboratory of Theoretical Physics of Gansu Province, School of Physical Science and Technology, Lanzhou University, Lanzhou 730000, China}
\author{Xin Zhang\footnote{Corresponding author}}
\email{zhangxin@mail.neu.edu.cn}
\affiliation{Department of Physics, College of Sciences, Northeastern University, Shenyang 110819, China}
\affiliation{Frontiers Science Center for Industrial Intelligence and Systems Optimization, Northeastern University, Shenyang 110819, China}
\affiliation{Key Laboratory of Data Analytics and Optimization for Smart Industry (Northeastern University), Ministry of Education, Shenyang 110819, China}



\begin{abstract}
Black hole shadow and gravitational lensing play important roles in testing gravitational theories in the strong field regime. As the first-order modifications from quantum gravity, the nonlocality can be manifested by black hole shadow and gravitational lensing. For example, the cut-off parameter introduced by nonlocality will affect the shape and size of the black hole shadow, and also affect the deflection angle of light rays. In this paper, we mainly investigate the effects of the nonlocality on the black hole shadow and the gravitational lensing for two types of rotating black holes in nonlocal gravity. It is found that the size of the black hole shadow decreases with the cut-off parameter since the nonlocality weakens the gravitational constant, and the shape of the shadow gets more deformed with the increase of the cut-off parameter. However, if the rotation parameter is small, the shape of the shadow is almost a circle even though the cut-off parameter approaches its maximum. The energy emission rate in both models is also studied. The results show that there is a peak for each curve and the peak decreases and shifts to the low frequency with the increase of the cut-off parameter. Besides, we also explore the shadow of both types of black holes surrounded by a nonmagnetized pressureless plasma which satisfies the separability condition. It is found that the plasma has a frequency-dependent dispersive effect on the size and shape of the black hole shadow. For the gravitational lensing, we find that the cut-off parameter of model A makes a positive contribution to the deflection angle, which can be compared with the contribution of the rotation parameter, while the cut-off parameter of model B makes a negative contribution which can be ignored. These results may be helpful for probing nonlocal gravity in future observations.
\end{abstract}






\maketitle

\section{Introduction}

As a prediction of Einstein's general relativity (GR), black holes are invisible objects usually believed to be formed in the gravitational collapse of massive astronomical objects.
It is well known that photons emitted from an illuminated source behind a black hole will form a two-dimensional dark zone in the observer's sky, which is the so-called black hole shadow. As an imprint of the black hole, the black hole shadow provides us with valuable information about
the black hole.
For example, one can extract the spin and charge of the black hole {\cite{Takahashi996,Tsukamoto043,Zakharov479,Hioki2009,Zakharov2014,Li2014}}
and constrain some new parameters introduced by modified gravities from the black hole shadow {\cite{Neves717,Moffat130,Amarilla2010,Amarilla2012,Amarilla2013}}. The shadow cast by a spherically symmetric black hole, which is a perfect circle, was first investigated by Synge~\cite{Synge1966} and then by Luminet~\cite{Luminet1979}. They also introduced formulas to calculate the angular radius and size of the shadow, respectively. The shadow cast by a Kerr black hole was first studied by Bardeen~\cite{Bardeen1973}. The shape of the shadow is deformed due to the dragging effect. Since then, the black hole shadow has been extensively investigated in the literature.
For instance, the black hole shadow and photon sphere in dynamically evolving spacetimes were investigated in ref.~\cite{Mishra2019}.
{ The effects of the cosmological constant on the black hole shadow were explored in refs.~\cite{Perlick104062,Haroon044015,Grenzebach2014,Li2020}}.
The shadow of a rotating regular black hole was explored in refs.~\cite{Amir2015,Abdujabbarov2016}. In ref.~\cite{Younsi2016}, the authors calculated the shadow of a model-independent parameterized axisymmetric black hole. The relations between the radius of the black hole shadow and the quasinormal modes in the eikonal limit were found in refs.~\cite{Konoplya2020,Liu2020,Jusufi2020a}. { Many other representative investigations on the black hole shadow have been carried out in, e.g., refs.~\cite{Atamurotov2013,Wei2013,Lu2014,Papnoi2014,Moffat2015a,Cunha2015,Cunha2017,Bisnovatyi-Kogan2018,Wei2019,Gralla2019,Kumar2020,Guo2020}.}

{
In addition to being captured, the escaped light rays will be bent by black holes, which is known as the gravitational lensing effect. As another powerful tool, the gravitational lensing also provides us abundant information about black holes, such as the position, mass, and angular momentum of black holes. Since the first detection of the deflection angle of light by the sun, the gravitational lensing has been extensively studied for black holes, wormholes, cosmic strings, and other objects by the geodesic method~\cite{Virbhadra084003,Virbhadra103004,Bozza103001,Perlick064017,Hoekstra606,Gallo083007,Nam1102}. In ref.~\cite{Gibbons235009}, Gibbons and Werner introduced an alternative method to calculate the weak deflection angle of light by a spherically symmetric black hole in the context of the optical geometry with the Gauss--Bonnet theorem. Then, Werner extended this method to the stationary black holes by using the Kerr--Randers optical geometry~\cite{Werner44}. With the Gibbons and Werner's methods, the weak deflection angle by different black holes in different gravitational theories was widely studied in refs.~\cite{Jusufi2016,Jusufi2017a,Jusufi2018a,Jusufi2018c,Arakida2018,Gyulchev2019,Javed2019,Kumar124024,Zhang015003}. Furthermore, the Gauss--Bonnet theorem has also been used to investigate the weak deflection angle of massive particles~\cite{Jusufi064017,Zonghai157}, charged particles~\cite{Li14226}, and for the finite receiver and light source~\cite{Ono2017,Ono2018,Crisnejo2019} in recent years.
}

In addition, for most astronomical situations, the influence of a plasma on light rays can be neglected but this is not true for the light rays in the radio frequency range. A well-known example is the effects of the solar corona, which is considered as a nonmagnetized pressureless plasma, on the time delay~\cite{Muhleman1966} and deflection angle when the light rays propagate near to the Sun~\cite{Muhleman1970}. Later on, Perlick performed a detailed investigation about the influence of a nonhomogeneous plasma on the light deflection in the Schwarzschild spacetime and in the equatorial plane of the Kerr spacetime~\cite{Perlick2000}. The effects of plasma on light propagation have attracted more and more attention since then. For instance, the influences of plasma on gravitational lensing by black holes and compact objects were investigated in refs.~\cite{Bisnovatyi-Kogan2010,Rogers2015,Morozova2013,Benavides-Gallego2018,Crisnejo2018,Bisnovatyi-Kogan2017}. The shadows of black holes and wormholes surrounded by plasma were investigated in refs.~\cite{Atamurotov2015,Perlick2015,Perlick2017,Yan2019}. For a review, see ref.~\cite{Bisnovatyi-Kogan2015}.

On the other hand, one of the most important motivations to study quantum gravity is the quest of ultraviolet complete gravitational theories, which may avoid spacetime singularity. Although we do not have a well-developed quantum gravity theory yet, there are many attempts, such as loop quantum gravity and string field theory. Almost all of these approaches include a common prediction that there should be an intrinsic extended structure in the spacetime~\cite{Das2008}, which is at the order of the Planck length. Such an extended structure indicates the nonlocality of the spacetime~\cite{Masood2016,Modesto2016a,Modesto2016b,Faizal2017}. Thus, it is believed that the first-order modifications from quantum gravity would be the nonlocal correction of GR~\cite{Elizalde1995,Modesto2011,Chicone2013,Joshi2250036}. As one of the current candidate quantum gravitational theories, nonlocal gravity
has been extensively studied in the literature. For instance, it was shown that nonlocal modifications offer a well-behaved ultraviolet complete quantum gravity which can be super-renormalizable~\cite{Moffat2010,Biswas2010,Modesto2012,Modesto2014}. The infrared nonlocal modifications of GR have been investigated in refs.~\cite{Maggiore2014,Narain2018,Kumar124040,Tian2019}, which can be used to explain the late-time cosmic acceleration. In refs.~\cite{Dirian2014,Dirian2015,Dirian2016}, the cosmological perturbation was analyzed and it was found that the IR nonlocally modified gravity not only performs well at the scale of the solar system as GR but can also provide a good fit to cosmic microwave background, baryon acoustic oscillation, and supernova data, as well as the $\Lambda$CDM model.

As special astronomical objects, black holes play an important role in probing the strong gravity regime where the quantum effects may dominate.
Thus, it is reasonable to study the black hole with quantum corrections. As the first-order corrections from quantum effects, it is interesting to consider the black hole in nonlocal gravity.
The spherically symmetric static black hole in nonlocal gravity has been considered in refs.~\cite{Modesto2011,Nicolini2012,Isi2013,Frolov2015,Knipfer2019}. However, most astrophysical black holes are formed with rotation. Thus, it is worth investigating the spinning black hole in nonlocal gravity. Additionally, to obtain deep insight into the characteristics of the nonlocally modified black hole and investigate the effects of the nonlocal modifications on the black hole, it is necessary to study the shadow {and weak deflection angle} of the black hole in this gravitational theory. Furthermore, an astronomical object, including black holes, is usually surrounded by plasma. Thus, it is also necessary to investigate the effects of the plasma on the black hole shadow.

This paper is organized as follows. In Section~\ref{nonlocal}, we give a brief review of the nonlocally modified gravitational theory and discuss two kinds of spherically symmetric black hole solutions. In Section~\ref{robh}, the rotating black hole solutions in the nonlocal gravity are presented and the null geodesics of the spacetime of these two types of rotating black holes are obtained. In Section~\ref{shadow}, we investigate the effects of the nonlocality on the black hole shadow. In Section~\ref{emr}, the energy emission rates of both models are investigated. The effects of a nonmagnetized pressureless plasma on the black hole shadow are analyzed in Section~\ref{plasma}. Section \ref{con} gives the conclusion.

\section{A Brief Review of the Nonlocal Gravity}~\label{nonlocal}
In this section, we first give a brief review of the nonlocal gravity and then discuss two kinds of spherically symmetric black hole solutions. One can see ref.~\cite{Nicolini2012} for more details and discussions. We begin with the following action~\cite{Moffat2010,Modesto2011,Nicolini2012}:
\begin{eqnarray}
S_{\text{tot}}=S_{\text{grav}}+S_{\text{matt}}, ~\label{action}
\end{eqnarray}
where
\begin{eqnarray}
S_{\text{grav}}=\frac{1}{16\pi G}\int d^4x\sqrt{-g}\mathcal{A}^{-2}(\square)R,
\end{eqnarray}
where $\mathcal{A}(\square)$ is an entire function of the dimensionless covariant d'Alembertian operator, i.e., $\square\equiv l^{2} g^{\mu\nu}\nabla_{\mu}\nabla_{\nu}$ with $l$ as a fundamental length scale of the theory. {$S_{\text{grav}}$ is originated from an ultraviolet complete quantum gravity where the Newton's constant is replaced by $G\mathcal{A}^{2}(\square)$~\cite{Modesto2011}. This theory not only is ultraviolet complete but also satisfies unitarity to all orders of perturbation theory. Furthermore, it should be stressed that here the d'Alembertian operator appears in the denominator, and thus the Lagrangian $\mathcal{A}^{-2}(\square)R$ is actually not a total derivative.} $S_{\text{matt}}$ is the action for the matter field. Varying the action (\ref{action}) with respect to the metric $g_{\mu\nu}$, one can obtain the following equations of motion~\cite{Moffat2010,Modesto2011,Nicolini2012}:
\begin{eqnarray}
\mathcal{A}^{-2}(\square)\left(R_{\mu\nu}-\frac{1}{2}g_{\mu\nu}R\right)=8\pi G T_{\mu\nu}.
\end{eqnarray}

By shifting the operator $\mathcal{A}^{-2}$ to the right hand side, the above equations can be rewritten as~\cite{Nicolini2012}
\begin{eqnarray}
R_{\mu\nu}-\frac{1}{2}g_{\mu\nu}R=8\pi G \mathcal{T}_{\mu\nu},~\label{eom2}
\end{eqnarray}
where the effective energy momentum tensor is defined as $\mathcal{T}_{\mu\nu}\equiv \mathcal{A}^{2}(\square)T_{\mu\nu}$, which is divergence free.

Assuming that the matter field is a pressureless static fluid at the origin, the component $T^0_0$ of its energy momentum tensor can be expressed as
\begin{eqnarray}
T^0_0=-\frac{M}{4\pi r^2}\delta(r),
\end{eqnarray}
where $M$ is the mass of the source and $\delta(r)$ is the usual delta function. Due to the spherically symmetric static fluid, the metric of the spacetime can be assumed as
\begin{eqnarray}
ds^2=-f(r)dt^2+\frac{1}{f(r)}dr^2+r^2(d\theta^2+\sin^2\theta d\phi^2), ~\label{smetric}
\end{eqnarray}
{where the metric components $g_{00}$ and $-g_{rr}^{-1}$ are assumed to be identical for simplicity}.

Inserting the metric (\ref{smetric}) into Equation~(\ref{eom2}), the metric component can be solved as
\begin{eqnarray}
f(r)=1-\frac{2\mathcal{G}(r)M}{r},
\end{eqnarray}
where the effective Newton’s constant is defined by
\begin{eqnarray}
\mathcal{G}(r)=-\frac{4\pi G}{M}\int dr r^2 \mathcal{T}^0_0,
\end{eqnarray}
which incorporates all the nonlocal effects and should reduce to Newton's constant for $r\gg l$.
Before solving $\mathcal{G}(r)$, one has to specify the form of $\mathcal{A}(\square)$. In this paper, we will take the following two models as examples.

\subsection{Model A}

We start with model A by assuming~\cite{Nicolini2012}
\begin{eqnarray}
\mathcal{A}(p^2)=\exp (-l p/2),
\end{eqnarray}
where $l$ is the {scale of the nonlocal} parameter. In the free-falling Cartesian-like coordinates, the component $\mathcal{T}^0_0$ of the effective energy momentum tensor can be conveniently calculated as
\begin{eqnarray}
\mathcal{T}^0_0&=&-M\mathcal{A}^{2}(\square)\delta(\vec{x})=-\frac{M}{(2\pi)^3}\int d^3p~\text{e}^{-l|\vec{p}|}\text{e}^{i\vec{x}\cdot \vec{p}} \nonumber\\
&=&-\frac{M}{\pi^2}\frac{l}{(\vec{x}^2+l^2)^2}.
\end{eqnarray}

Then, the effective Newton’s constant can be expressed as
\begin{eqnarray}
\mathcal{G}(r)\!\!\!&=&\!\!\!-\frac{4\pi G}{M}\!\!\int\!\! dr r^2 \mathcal{T}^0_0
\!\!=\!\!\frac{2 G}{\pi}\bigg(\!\!\arctan(r/l)\!-\!\frac{r/l}{1\!+\!(r/l)^2}\!\!\bigg). ~~~~~
~\label{mofa}
\end{eqnarray}

{It is interesting to find that the solution of this model not only matches those derived in the context of asymptotically safe gravity~\cite{Platania470} but also the noncommutative geometry inspired black holes except for $l$ being replaced by $\sqrt{\pi\theta}$~\cite{Nicolini547}.}

For $r\gg l$, $\mathcal{G}(r)$ can be expanded as
\begin{eqnarray}
\mathcal{G}(r)\approx G\left[1-\frac{4}{\pi}\left(\frac{l}{r}\right)+\frac{8}{3\pi}\left(\frac{l}{r}\right)^3\right],
\end{eqnarray}
which indicates that the effective Newton’s constant $\mathcal{G}(r)$ reduces to Newton's constant $G$ at a large distance and the metric (\ref{smetric}) tends to the Schwarzschild spacetime. On the other hand, one can expand $\mathcal{G}(r)$ for $r\ll l$ as
\begin{eqnarray}
\mathcal{G}(r)\approx \frac{4}{3\pi}\left(\frac{r}{l}\right)^3 G,
\end{eqnarray}
which indicates that this theory is asymptotically {safe}~\cite{Platania188}.

\begin{figure}[htb]
{\includegraphics[width=6cm]{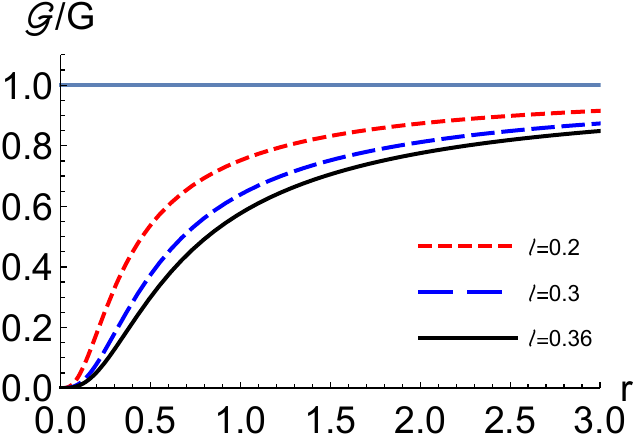}
\caption{Shapes of the function $\mathcal{G}(r)/G$.}
~\label{gA}}
\end{figure}

Figure \ref{gA} shows the shapes of $\mathcal{G}(r)/G$ for different $l$. It is obvious that the nonlocally modified gravitational theory reduces to GR in the infrared limit and is asymptotically {safe} in the ultraviolet limit.

Let us then check the regularity of the spacetime of model A by calculating the following curvature invariants
\begin{eqnarray}
R\!\!\!&=&\!\!\!\frac{32 G l^3 M}{\pi  \left(l^2+r^2\right)^3}, \\
R_{\mu\nu}R^{\mu\nu}\!\!\!&=&\!\!\!\frac{4096 G^2 l^2 M^2 \left(l^4+r^4\right)}{\pi ^2 \left(l^2+r^2\right)^6}, \\
R_{\mu\nu\rho\sigma}R^{\mu\nu\rho\sigma}\!\!\!&=&\!\!\!\frac{64 G^2 M^2}{\pi ^2 r^6 \left(l^2+r^2\right)^6}\Big(3 \left(l^2+r^2\right)^6 \tan ^{-1}\left(\frac{r}{l}\right)^2 \nonumber\\
\!\!\!&+&\!\!\!l^2 r^2 \left(3 l^4+4 l^2 r^2+5 r^4\right) \left(l^4+4 l^2 r^2+7 r^4\right) \nonumber\\
\!\!\!&-&\!\!\!2 l r \!\!\left(3 l^4\!+\!8 l^2 r^2\!+\!9 r^4\right) \left(l^2\!+\!r^2\right)^3 \tan ^{-1}\!\!\left(\frac{r}{l}\right)\!\!\Big) . ~~~ \nonumber\\
\end{eqnarray}
It is obvious that the scalars $R$ and $R_{\mu\nu}R^{\mu\nu}$ are finite everywhere but the regularity of $R_{\mu\nu\rho\sigma}R^{\mu\nu\rho\sigma}$ is implicit. For $r\ll l$, the Kretschmann scalar can be expanded as
\begin{eqnarray}
R_{\mu\nu\rho\sigma}R^{\mu\nu\rho\sigma}\approx \frac{512 G^2 M^2}{3 \pi ^2 l^6}-\frac{1024 G^2 M^2 r^2}{\pi ^2 l^8},
\end{eqnarray}
which is regular as desired. Thus, one can conclude that the spacetime of model A is free of singularity. {However, it should be pointed out that this model had been excluded based on the validity of the principle of least action at large distances~\cite{Benjamin01216}.}

\subsection{Model B}

As another example, inspired by the generalized uncertainty principle effects in the gravitational field~\cite{Isi2013}, we consider the following function~\cite{Nicolini2012}:
\begin{eqnarray}
\mathcal{A}(p^2)=(1+\beta p^2)^{-\frac{1}{2}},
\end{eqnarray}
where $\beta$ is the {scale of the nonlocal} parameter. Then, the effective energy momentum tensor can be calculated as
\begin{eqnarray}
\mathcal{T}^0_0=-M\mathcal{A}^{2}(\square)\delta(\vec{x})=-\frac{M}{\beta}\frac{\text{e}^{-|\vec{x}|/\sqrt{\beta}}}{4\pi|\vec{x}|}.
\end{eqnarray}

\begin{figure}[htb]
{\includegraphics[width=6cm]{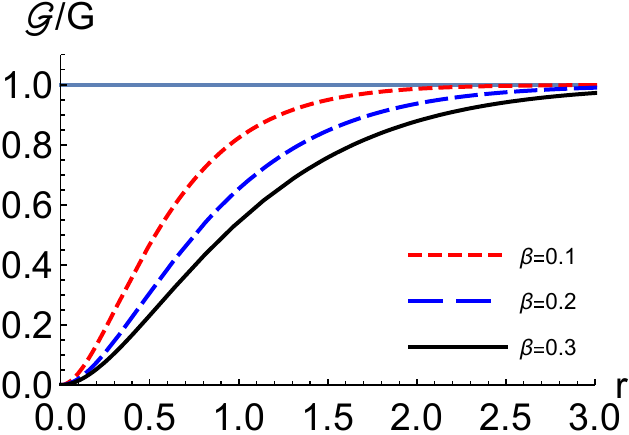}
\caption{Shapes of the function $\mathcal{G}(r)/G$.}
~\label{gB}}
\end{figure}

The effective Newton’s constant reads
\begin{eqnarray}
\mathcal{G}(r)=-\frac{4\pi G}{M}\int dr r^2 \mathcal{T}^0_0=G\gamma(2;r/\sqrt{\beta}),
\end{eqnarray}
where $\gamma(s;x)=\int^x_0 dt~t^{s-1}\text{e}^{-t}$ is the incomplete gamma function. Figure \ref{gB} shows that the effective Newton’s constant is zero at the origin and approaches Newton's constant at a large distance, which indicates that model B is also asymptotically {safe} in the ultraviolet limit and reduces to Einstein's general gravity in the infrared limit. However, the singularity at the origin is not removed, which can be seen from the short scale behavior of the Kretschmann scalar
\begin{eqnarray}
\lim_{r\rightarrow 0} R_{\mu\nu\rho\sigma}R^{\mu\nu\rho\sigma}\sim\frac{8 G^2 M^2}{\beta ^2 r^2}-\frac{16 G^2 M^2}{\beta ^{5/2} r}.
\end{eqnarray}

\section{Spinning Nonlocal Black Hole and Circular Photon Orbits}~\label{robh}
The Newman--Janis algorithm (NJA) is a generating method for constructing rotating black hole solutions from static ones~\cite{Newman1965}. This method works well in producing Kerr black holes from Schwarzschild black holes~\cite{Newman1965} and Kerr--Newman black holes from Reissner--Nordstr$\ddot{\text{o}}$m black holes in general relativity~\cite{Couch1965}. However, for other spherically symmetric black holes in modified gravitational theories, their rotating counterparts obtained by using the NJA will introduce additional sources~\cite{Johannsen2011,Bambi2013,Ghosh2015,Jusufi2020b}. We derive the rotating black hole solution in the nonlocal gravity by using the NJA without complexification~\cite{Azreg2014}, which has been successfully used to generate imperfect fluid stationary and rotating solutions from spherically symmetric ones.
In this section, we first present the rotating counterpart of the spherically symmetric static solution in the nonlocal gravity and then investigate the photon orbits in this background. In the Boyer--Lindquist coordinates, the resulting rotating solution reads as
\begin{eqnarray}~\label{rsol}
ds^2&=&-\left(\frac{\Delta-a^2\sin^2\theta}{\rho^2}\right)dt^2+\frac{\rho^2}{\Delta}dr^2+\rho^2 d\theta^2 \nonumber\\
&-&2a\sin^2\theta\left(1-\frac{\Delta-a^2\sin^2\theta}{\rho^2}\right)dtd\phi \\
&+&\sin^2\theta\left[\rho^2+a^2\sin^2\theta\left(2-\frac{\Delta-a^2\sin^2\theta}{\rho^2}\right)\right]d\phi^2, \nonumber
\end{eqnarray}
where
\begin{eqnarray}
\Delta=r^2+a^2-2M\mathcal{G}(r)r, \quad \rho^2=r^2+a^2\cos^2\theta, ~\label{kerrm}
\end{eqnarray}
with $a$ as the rotation parameter. Different from the case of a spherically symmetric static solution, the rotating solution (\ref{rsol}) needs some fluid or other nonvacuum sources, which is a complex question worth studying in future work. Here, we will investigate the effects of the {nonlocal} parameter on the lensing of light in this geometry.

The event horizon of this rotating black hole is determined by
\begin{eqnarray}
g^{\mu\nu}\partial_{\mu}r\partial_{\nu}r=g^{rr}=\Delta(r)=0.
\end{eqnarray}

\begin{figure*}[htb]
\begin{center}
\subfigure[]  {\label{rpa}
\includegraphics[width=6cm]{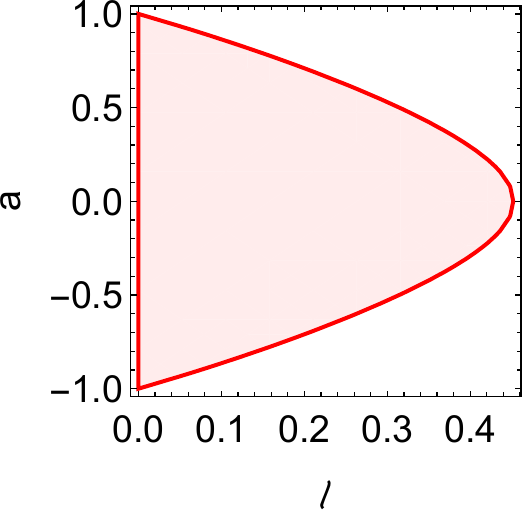}}
\subfigure[]  {\label{rpb}
\includegraphics[width=6cm]{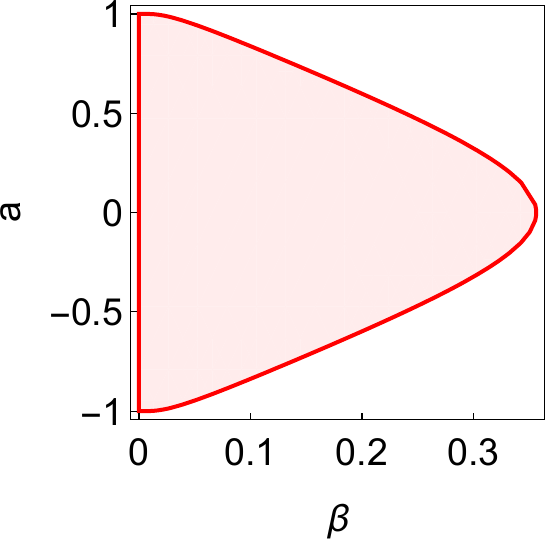}}
\end{center}
\caption{The parameter spaces determined by the existence condition of black hole event horizon. Panels (\textbf{a}) and (\textbf{b}) show the cases of models A and B, respectively.}
\label{rp}
\end{figure*}

The feasible ranges of the rotation parameter $a$ and the {nonlocal} parameters $l$ and $\beta$ are determined by the existence condition of the horizons $\Delta(r_{\text{c}})\leq0$, where $r_c$ is given by the zeros of $\frac{d\Delta(r_{\text{c}})}{dr}=0$. In Figure~\ref{rp}, the parameter spaces $(a,l)$ and $(a,\beta)$ are shown as the light pink regions. In the following, we adimensionalize all quantities with the mass of the black hole and set Newton's constant to one.

\subsection{Equations of Null Geodesic Motion}

The motion of a photon in the black hole spacetime can be described by the Lagrangian
\begin{eqnarray}
\mathcal{L}=\frac{1}{2}g_{\mu\nu}\dot{x}^{\mu}\dot{x}^{\nu},
\end{eqnarray}
where $\dot{x}^{\mu}=u^{\mu}=dx^{\mu}/d\lambda$, $u^{\mu}$ is the photon's 4-velocity, and $\lambda$ is the affine parameter. Then, the energy and angular momentum are given by
\begin{eqnarray}
E&=&-p_t=-\frac{\partial \mathcal{L}}{\partial \dot{t}}=-g_{\phi t}\dot{\phi}-g_{tt}\dot{t}, \\
L&=&p_{\phi}=\frac{\partial \mathcal{L}}{\partial \dot{\phi}}=g_{\phi \phi}\dot{\phi}+g_{\phi t}\dot{t}.
\end{eqnarray}

Due yo the stationary and axisymmetric properties of the spacetime, both the energy and angular momentum are conserved.

The corresponding Hamilton--Jacobi equation is given by {\cite{Chandrasekhar1983}}
\begin{eqnarray}
\frac{\partial S}{\partial \lambda}=-\frac{1}{2}g^{\mu\nu}\frac{\partial S}{\partial x^{\mu}}\frac{\partial S}{\partial x^{\nu}}. ~\label{Seom}
\end{eqnarray}

Considering the symmetry of the spacetime and the separability of the solution, the Jacobi action $S$ for photons can be assumed as
\begin{eqnarray}
S=-Et+L\phi+S_r(r)+S_{\theta}(\theta), ~\label{Saction}
\end{eqnarray}
where $S_r(r)$ and $S_{\theta}(\theta)$ are only functions of $r$ and $\theta$, respectively. Substituting Equation~(\ref{Saction}) into Equation~(\ref{Seom}), one can obtain {\cite{Bardeen347,Cunningham237,Cunningham788,Chandrasekhar1983}}
\begin{eqnarray}
\rho^2 \frac{\partial t}{\partial \lambda}\!\!&=&\!\!a(L\!-\!a E \sin^2\theta)\!+\!\frac{r^2\!+\!a^2}{\Delta}[E(r^2\!+\!a^2)\!-\!a L],~~~\\
\rho^2 \frac{\partial \phi}{\partial \lambda}\!\!&=&\!\!(L \csc^2\theta-a E)+\frac{a}{\Delta}[E(r^2+a^2)-a L], \\
\rho^2 \frac{\partial r}{\partial \lambda}\!\!&=&\!\!\pm\sqrt{\mathcal{R}}, ~\label{rm} \\
\rho^2 \frac{\partial \theta}{\partial \lambda}\!\!&=&\!\!\pm\sqrt{\Theta},~\label{them}
\end{eqnarray}
where
\begin{eqnarray}
\mathcal{R}&=&[E(r^2+a^2)-a L]^2-\Delta[\mathcal{Q}+(L-a E)^2], \\
\Theta&=&\mathcal{Q}+\cos^2\theta(a^2E^2-L^2\csc^2\theta),
\end{eqnarray}
and $\mathcal{Q}$ is the Carter constant {\cite{Carter280}}. These equations determine the null geodesics of the spacetime of the rotating nonlocally modified black hole. The plus and minus signs in the radial Equation (\ref{rm}) correspond to the outgoing and ingoing photons, respectively. The plus and minus signs in Equation~(\ref{them}) correspond to the photons moving to the north ($\theta=0$) and south ($\theta=\pi$) poles, respectively.

\subsection{Unstable Circular Photon Orbits}~\label{sectionB}
The silhouette of the black hole is formed by the unstable circular photon orbits with constant $r$, which should satisfy
\begin{eqnarray}
\mathcal{R}(r)=0, \quad \frac{d\mathcal{R}}{dr}=0.
\end{eqnarray}

From the above two conditions, one can obtain {\cite{Bardeen347,Cunningham237,Chandrasekhar1983}}
\begin{eqnarray}
\xi&\equiv& \frac{L}{E}=\frac{-4r \Delta+(a^2+r^2)\Delta'}{a\Delta'}, \\
\eta&\equiv&\frac{\mathcal{Q}}{E^2}=\frac{r^2\big(16\Delta(a^2-\Delta)+8r\Delta\Delta'-r^2\Delta'^2\big)}{a^2\Delta'^2},
\end{eqnarray}
where $\xi$ and $\eta$ are two {constants}, which are important for determining the contour of the black hole shadow. The prime denotes the derivative with respect to the radial \mbox{coordinate $r$.}

\section{Shadow of Black Hole}~\label{shadow}
In general, if there is a black hole between a light source and an observer, the photons with small orbital angular momentum emitted from the light source will be absorbed by the black hole and form a dark zone in the observer's sky, which is known as the black hole shadow. {To investigate the shadow viewed by a static observer located at infinite distance, we use the following basis vectors~\cite{Bardeen347,Chandrasekhar1983}:

\begin{eqnarray}
\hat{e}_{(t)}&=&\sqrt{\frac{A}{\rho^2\Delta}}\left(\partial_t+\frac{2Mar\mathcal{G}}{A}\partial_{\phi}\right) , \\
\hat{e}_{(r)}&=&\sqrt{\frac{\Delta}{\rho^2}}\partial_r ,\\
\hat{e}_{(\theta)}&=&\frac{1}{\sqrt{\rho^2}}\partial_{\theta} ,\\
\hat{e}_{(\phi)}&=&\frac{1}{\sin\theta}\sqrt{\frac{\rho^2}{A}}\partial_{\phi},
\end{eqnarray}
with $A\equiv \left(r^2+a^2\right)^2-a^2\Delta \sin ^2\theta$.
The projected four-momentum of the photon with respect to the observer is given by
\begin{eqnarray}
p^{(t)}&=&-p_{\mu}\hat{e}^{\mu}_{(t)}=-\sqrt{\frac{A}{\rho^2\Delta}}\left(p_t+\frac{2Mar\mathcal{G}}{A}p_{\phi}\right) \nonumber\\
&=&-\sqrt{\frac{A}{\rho^2\Delta}}\left(-E+\frac{2Mar\mathcal{G}}{A}L\right), \\
p^{(r)}&=&p_{\mu}\hat{e}^{\mu}_{(r)}=\sqrt{\frac{\Delta}{\rho^2}}p_r=\sqrt{\frac{\mathcal{R}}{\rho^2\Delta}}, \\
p^{(\theta)}&=&p_{\mu}\hat{e}^{\mu}_{(\theta)}=\frac{p_{\theta}}{\sqrt{\rho^2}}=\sqrt{\frac{\Theta}{\rho^2}}, \\
p^{(\phi)}&=&p_{\mu}\hat{e}^{\mu}_{(\phi)}=\frac{1}{\sin\theta}\sqrt{\frac{\rho^2}{A}}p_{\phi}=\frac{1}{\sin\theta}\sqrt{\frac{\rho^2}{A}}L.
\end{eqnarray}

Then, the celestial coordinates on the plane of the observer's sky can be calculated \mbox{as~\cite{Bardeen347,Cunningham237,Cunningham788,Chandrasekhar1983}}
\begin{eqnarray}
x\!\!&=&\!\!\lim_{r\rightarrow\infty}\left(\frac{r p^{(\phi)}}{p^{(t)}}\right)=-\xi\csc\theta_o, \\
y\!\!&=&\!\!\lim_{r\rightarrow\infty}\left(\frac{r p^{(\theta)}}{p^{(t)}}\right)=\pm\sqrt{\eta+a^2\cos^2\theta_o-\xi^2\cot^2\theta_o},~~~
\end{eqnarray}
}where $\theta_o$ is the angle of inclination between the rotation axis of the black hole and the line of sight of the observer. We mainly focus on the influence of the nonlocal correction on the shadow. Thus, for simplicity, we assume that the observer is located at the equatorial plane of the black hole, i.e., $\theta_o=\frac{\pi}{2}$. Then, the celestial coordinates reduce to
\begin{eqnarray}
x&=&-\xi, \\
y&=&\pm\sqrt{\eta}.
\end{eqnarray}

\begin{figure*}[htb]
\begin{center}
\subfigure[~$l=0.01$]  {\label{saa}
\includegraphics[width=6cm]{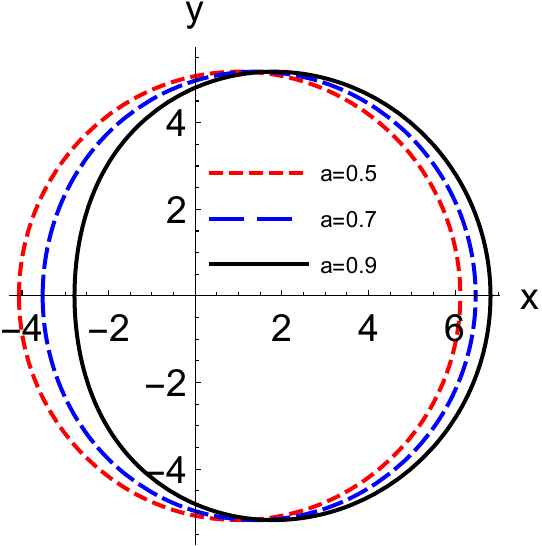}}
\subfigure[~$a=0.6$]  {\label{sal}
\includegraphics[width=5.53cm]{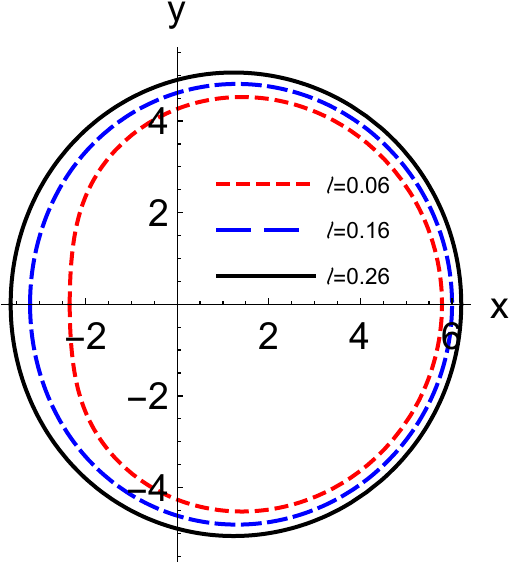}}
\end{center}
\caption{The shadow cast by the rotating black hole in model A. {Panels (\textbf{a}) and (\textbf{b}) show the effects of the rotation parameter $a$ and the nonlocal parameter $l$ on the black hole shadow, respectively.} Here, the black hole is located at the origin of coordinates with the angle of inclination $\theta_o=\frac{\pi}{2}$.}
\label{sa}
\end{figure*}

\begin{figure*}[htb]
\begin{center}
\subfigure[~$\beta=0.01$]  {\label{shba}
\includegraphics[width=6cm]{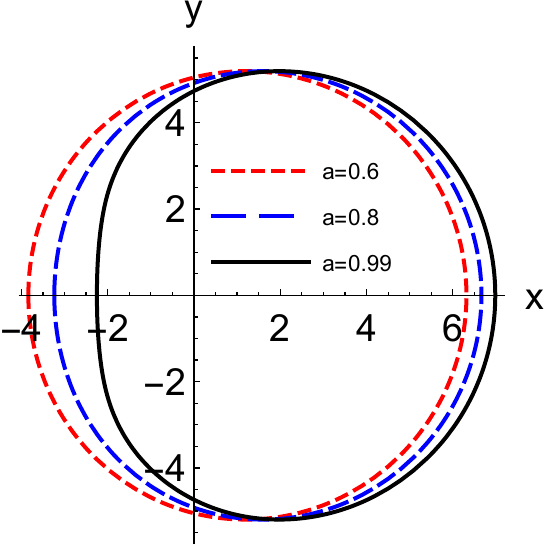}}
\subfigure[~$a=0.01$]  {\label{shbb}
\includegraphics[width=5.53cm]{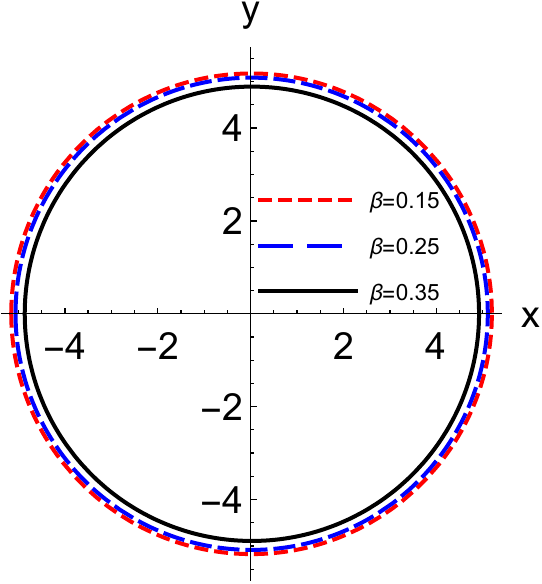}}
\end{center}
\caption{The shadow cast by the rotating black hole in model B. {Panels (\textbf{a}) and (\textbf{b}) show the effects of the rotation parameter $a$ and the nonlocal parameter $\beta$ on the black hole shadow, respectively.} Here, the black hole is located at the origin of coordinates with the angle of inclination $\theta_o=\frac{\pi}{2}$.}
\label{shb}
\end{figure*}

Figures \ref{sa} and \ref{shb} present the shadows of the rotating black holes of model A and model B with different rotation parameters and {nonlocal} parameters, respectively. It is obvious that the silhouette is more deformed with the increase in the rotation parameter $a$ for fixed {nonlocal} parameter $l$ or $\beta$ while the size of it almost remains the same except shifting to the right; see Figures~\ref{sa}a and \ref{shb}a.
Additionally, Figures~\ref{sa}b and \ref{shb}b show that the silhouette is also more deformed with the increase in $l$ or $\beta$ for fixed $a$. However, the deformation is suppressed when the rotation parameter is small even if the {nonlocal} parameter approaches its maximum; see Figure~\ref{shb}b.

In order to determine the important astronomical information of the black hole, it is necessary to construct some astronomical observables. Following Hioki and Maeda's proposals~\cite{Hioki2009}, two observables are introduced to characterize the apparent shape of the shadow, i.e., the radius $R_s$ and the distortion parameter $\delta_s$. The observable $R_s$ is the radius of a reference circle passing through three points: the top point ($x_t$, $y_t$), the bottom one ($x_b$, $y_b$), and the rightmost one ($x_r$, 0) of the shadow. The distortion parameter $\delta_s$ measures the deformation of the shadow with respect to the reference circle and is defined as $\delta_s=D/R_s$, where $D$ is the difference between the leftmost points of the reference circle and of \mbox{the shadow.}

After a simple algebra calculation, these two observables can be expressed as {\cite{Hioki2009,Wei2013,Abdujabbarov2016}}
\begin{eqnarray}
R_s&=&\frac{(x_t-x_r)^2+y_t^2}{2(x_r-x_t)},~\label{rd} \\
\delta_s&=&\frac{x_1-\tilde{x}_r}{R_s},
\end{eqnarray}
where $(\tilde{x}_r,0)$ and $(x_1,0)$ are the points where the reference circle and the contour of the shadow intersect the horizontal axis at the opposite side of $(x_r,0)$, respectively.

\begin{figure*}[htb]
\begin{center}
\subfigure[$~$]  {\label{Rsa}
\includegraphics[width=6.5cm]{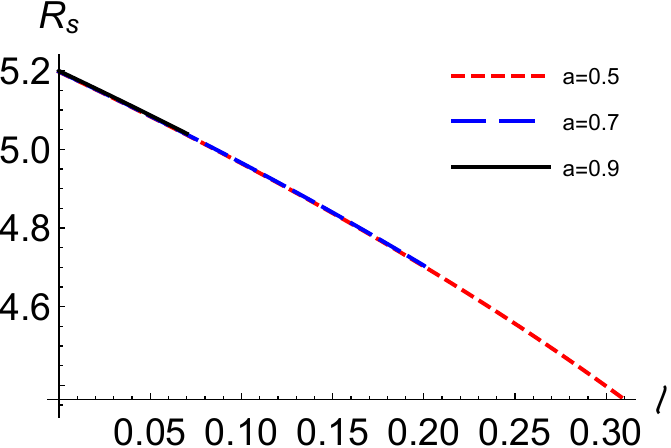}}
\subfigure[$~$]  {\label{dsa}
\includegraphics[width=6.5cm]{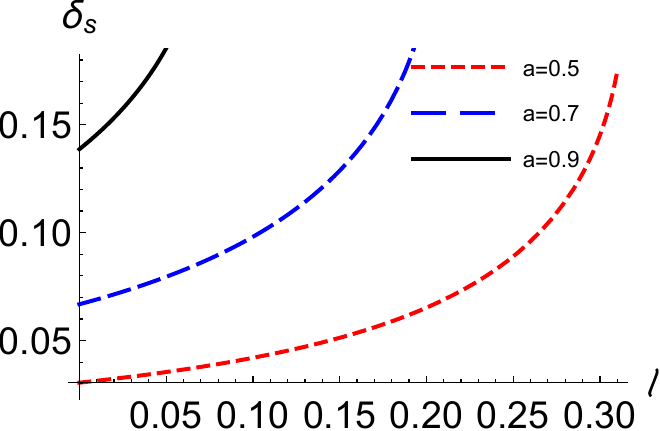}}
\end{center}
\caption{{Observables $R_s$ and $\delta_s$ as functions of the nonlocal parameter $l$ in model A, as shown in panels (\textbf{a}) and (\textbf{b}), respectively.} Here, the rotating black hole is located at the origin of coordinates with the angle of inclination $\theta_o=\frac{\pi}{2}$.}

\label{oba}
\end{figure*}

\begin{figure*}[htb]
\begin{center}
\subfigure[$~$]  {\label{Rsb}
\includegraphics[width=6.5cm]{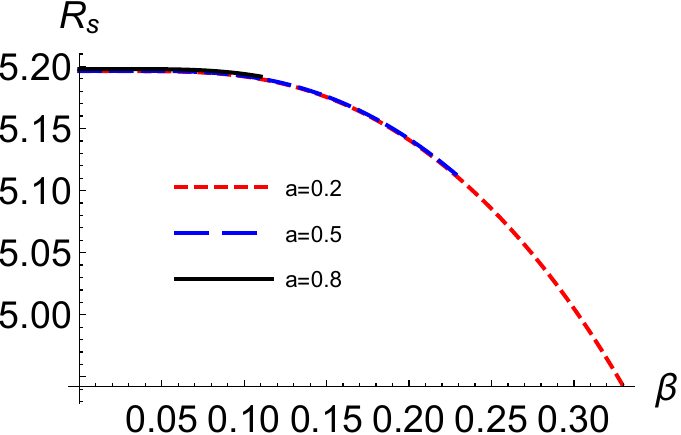}}
\subfigure[$~$]  {\label{dsb}
\includegraphics[width=6.5cm]{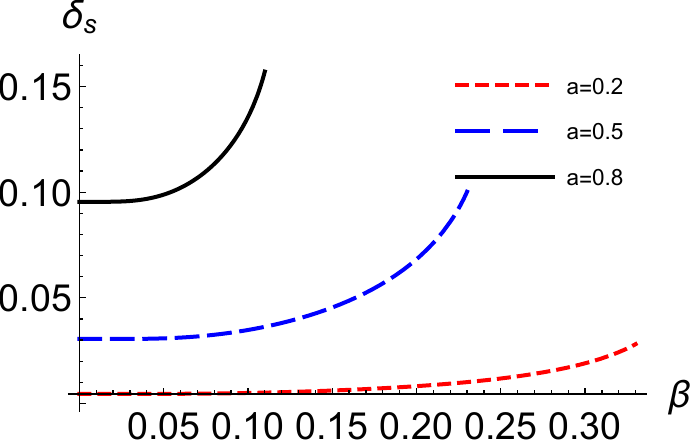}}
\end{center}
\caption{{Observables $R_s$ and $\delta_s$ as functions of the nonlocal parameter $\beta$ in model B, as shown in panels (\textbf{a}) and (\textbf{b}), respectively.} Here, the rotating black hole is located at the origin of coordinates with the angle of inclination $\theta_o=\frac{\pi}{2}$.}

\label{obb}
\end{figure*}

Figures \ref{oba} and \ref{obb} show the observables $R_s$ and $\delta_s$ of the black hole shadow of model A and model B as a function of the {nonlocal} parameter $l$ or $\beta$ for the different values of the rotation parameter $a$, respectively. It is obvious that the observable $R_s$ decreases with $l$ or $\beta$.
The reason is that with the increasing $l$ or $\beta$, the effective gravitational constant will decrease (see Figures~\ref{gA} and \ref{gB}), and hence the gravitational force acting on photons becomes weaker.
In addition, the values of $R_s$ are approximately equal for different $a$, which makes the curves indistinguishable in Figures \ref{oba}a and \ref{obb}a. This is because the rotation parameter $a$ only deforms the shape of the black hole shadow and has almost no influence on the size of the black hole; see Tables \ref{tableoba} and \ref{tableobb}. The observable $\delta_s$ gives the distortion of the shape of the shadow from the reference circle, which increases with $l$ or $\beta$. Furthermore, the shadow becomes more deformed with larger $a$ for fixed $l$ or $\beta$; see Figures \ref{oba}b and \ref{obb}b or Tables \ref{tableoba} and \ref{tableobb}.

{Before ending this section, we give a rough estimation on the nonlocal parameters $l$ and $\beta$ by using the black hole shadow of M87*. According to the data provided by the EHT~\cite{EHTL1}, the mass of M87* is $M_{\text{M87}^*}=(6.5\pm 0.9) \times 10^9 M_{\odot}$, the distance of M87* from the Earth is $D=16.8^{+0.8}_{-0.7}$Mpc, and the angular diameter of the shadow is $\theta_{M87*}=42\pm 3 \mu$as. Then, the size of the shadow in units of mass can be calculated as $d_{M87*}=(11\pm1.5)~M$~\cite{Bambi044057}. By comparing $d_{M87*}$ with the theoretical shadow diameter $d=2R_s$, we find that the constraint on the nonlocal parameter $l$ is $0\leqslant l\leqslant 0.18$, while there is no constraint on $\beta$ since the radius of the black hole shadow always remains within the radius of M87* for the whole range $0\leqslant \beta \leqslant 0.35$.
}

\begin{table}
    \begin{center}
        \begin{tabular}{| c | c | c | c | c | c | c |}
            \hline
        &\multicolumn{2}{c|}{$a=0.5$}  &\multicolumn{2}{c|}{$a=0.7$} &\multicolumn{2}{c|}{$a=0.9$} \\
            \hline
    $\beta$ & $R_s$  & $\delta_s$  & $R_s$ & $\delta_s$
    & $R_s$ & $\delta_s$ \\
            \hline
    0.00    & 5.196  & 0.030  & 5.197  & 0.066 & 5.198  & 0.138 \\
    0.03    & 5.129  & 0.033  & 5.130  & 0.073 & 5.132  & 0.162 \\
    0.05    & 5.083  & 0.035  & 5.084  & 0.079 & 5.086  & 0.185 \\
    0.13    & 4.889  & 0.047  & 4.890  & 0.114 & ---  & --- \\
    0.15    & 4.837  & 0.051  & 4.839   & 0.128 & ---  & --- \\
    0.18    & 4.757  & 0.058  & 4.759  & 0.161 & ---  & --- \\
    0.23    & 4.616  & 0.077  & ---  & --- & ---  & --- \\	
            \hline
        \end{tabular}
        \caption{Observables of the rotating black hole in model A for different values of the parameter $l$.}\label{tableoba}
    \end{center}
\end{table}

\begin{table}
    \begin{center}
        \begin{tabular}{| c | c | c | c | c | c | c |}
            \hline
        &\multicolumn{2}{c|}{$a=0.2$}  &\multicolumn{2}{c|}{$a=0.5$}
        &\multicolumn{2}{c|}{$a=0.8$} \\
            \hline
    $\beta$ & $R_s$ & $\delta_s$  & $R_s$ & $\delta_s$
    & $R_s$ & $\delta_s$ \\
            \hline
    0.00    & 5.1961  & 0.0045  & 5.1964  & 0.0305 & 5.1979  & 0.0954 \\
    0.05    & 5.1960  & 0.0045  & 5.1963  & 0.0307 & 5.1978  & 0.0988 \\
    0.10    & 5.1919  & 0.0049  & 5.1923  & 0.0347 & 5.1939  & 0.1354 \\
    0.15    & 5.1753  & 0.0060  & 5.1758  & 0.0453 & ---  & --- \\
    0.20    & 5.1408  & 0.0081  & 5.1416   & 0.0682 & ---  & --- \\
    0.25    & 5.0853  & 0.0117  & ---  & --- & ---  & --- \\
    0.30    & 5.0051  & 0.0189  & ---  & --- & ---  & --- \\	
            \hline
        \end{tabular}
        \caption{Observables of the rotating black hole in model B for different values of the parameter $\beta$.}\label{tableobb}
    \end{center}
\end{table}

\section{Energy Emission Rate}~\label{emr}
In this section, we study the energy emission rate for the rotating black holes in both models. The expression of the energy emission rate can be read as~\cite{Mashhoon1973,Wei2013}
\begin{eqnarray}
\frac{d^2E(\omega)}{d\omega dt}=\frac{2\pi^3R_s^2}{\text{e}^{\omega/T_H}-1}\omega^3,
\end{eqnarray}
where $R_s$ is the radius of the shadow given in Equation~(\ref{rd}) and $\omega$ is the frequency of the photon. The Hawking temperatures for model A and model B can be calculated as {\cite{Parikh5042,Padmanabhan49}}

\begin{figure*}[htb]
\begin{center}
\subfigure[~$a=0.6$]  {\label{Eal}
\includegraphics[width=6.5cm]{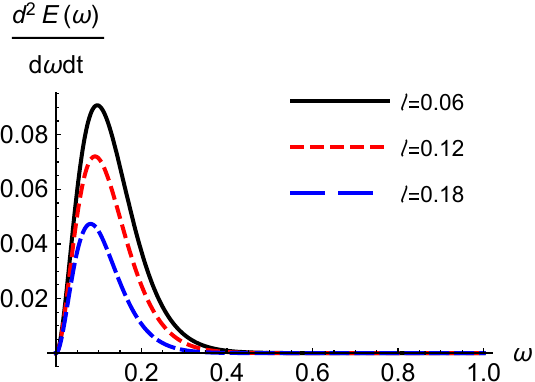}}
\subfigure[~$l=0.2$]  {\label{Eaa}
\includegraphics[width=6.5cm]{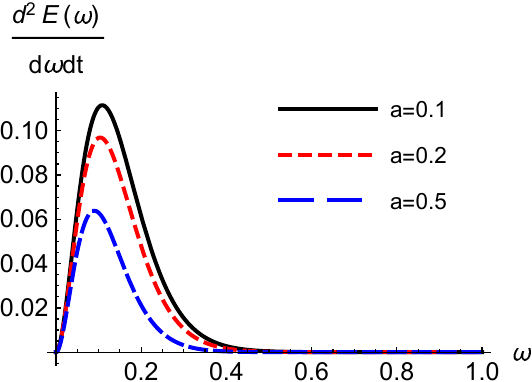}}
\end{center}
\caption{{The energy emission rate $\frac{d^2E(\omega)}{d\omega dt}$ in model A. Panels (\textbf{a}) and (\textbf{b}) show the effects of the nonlocal parameter $l$ and the rotation parameter $a$, respectively.}}
\label{Ea}
\end{figure*}

\begin{figure*}[htb]
\begin{center}
\subfigure[~$a=0.2$]  {\label{Eb1}
\includegraphics[width=6.5cm]{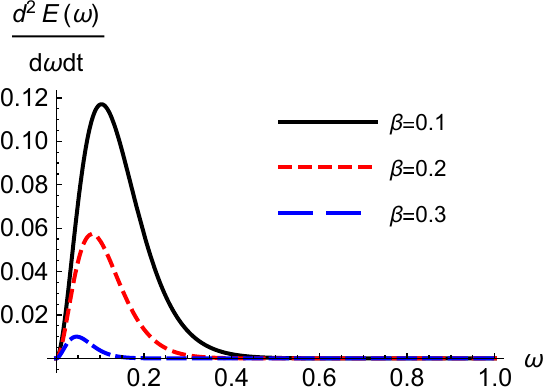}}
\subfigure[~$\beta=0.01$]  {\label{Eb5}
\includegraphics[width=6.5cm]{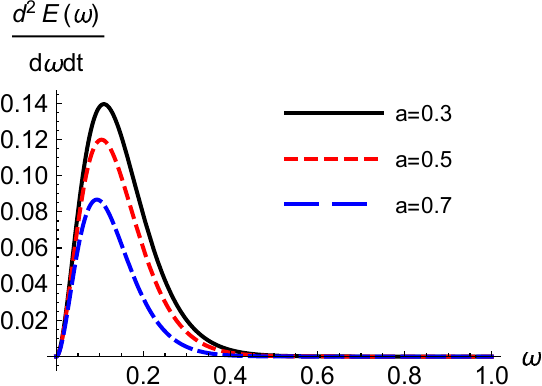}}
\end{center}
\caption{{The energy emission rate $\frac{d^2E(\omega)}{d\omega dt}$ in model B. Panels (\textbf{a}) and (\textbf{b}) show the effects of the nonlocal parameter $\beta$ and the rotation parameter $a$, respectively.}}
\label{Eb}
\end{figure*}

\begin{widetext}
\begin{eqnarray}
T_H^{(A)}\!\!\!\!&=&\!\!\!\!\frac{M \left(a^2 l^3 r_+-l r_+^3 \left(a^2+l^2\right)+\left(a-r_+\right) \left(a+r_+\right) \left(l^2+r_+^2\right){}^2 \tan ^{-1}\left(r_+/l\right)-3 l r_+^5\right)}{\pi ^2 \left(a^2+r_+^2\right){}^2 \left(l^2+r_+^2\right){}^2}, ~~\\
T_H^{(B)}\!\!\!\!&=&\!\!\!\!\frac{M \left(2\beta \left(a-r_+\right) \left(a+r_+\right) \left(\gamma \left(2,r_+/\sqrt{\beta }\right)-1\right)-2 r_+^2 \left(a^2+r_+^2\right) e^{-r_+/\sqrt{\beta }}\right)}{4 \pi\beta  \left(a^2+r_+^2\right)^2},
\end{eqnarray}
\end{widetext}
where $r_+$ is the outer horizon of the black hole.
In Figures~\ref{Ea} and \ref{Eb}, we plot $d^2E(\omega)/d\omega dt$ against frequency $\omega$ for both models. It is clear that there exists a peak for each curve and the peak decreases and shifts to the low frequency with increasing $l$ or $\beta$ for fixed $a$. Additionally, the peak also decreases and shifts to the low frequency with increasing $a$ for fixed $l$ or $\beta$.

\section{The Case with Plasma}~\label{plasma}
In this section, we investigate the shadow of the nonlocally modified black hole in the presence of the plasma. Let us consider a nonmagnetized pressureless plasma with \mbox{plasma frequency}
\begin{eqnarray}
\omega_p(x)^2=\frac{4\pi e^2N_e(x)}{m_e},
\end{eqnarray}
where $e$ and $m_e$ denote the charge and mass of an electron, respectively, and $N_e(x)$ stands for the number density of electrons in the plasma.
Then, photons propagating in this plasma can be given by the modified Hamiltonian
{\cite{Bisnovatyi-Kogan2010,Rogers2015,Perlick2017}}
\begin{eqnarray}
H(x,p)=\frac{1}{2}\left[g^{\mu\nu}(x)p_{\mu}p_{\nu}+\omega_p(x)^2\right]=0,~\label{Ham}
\end{eqnarray}
where $p_{\mu}$ is the 4-momentum of the photon. Provided the plasma frequency $\omega_p$ is only a function of $r$ and $\theta$, the energy and angular momentum of the photon are still constants since $\partial_t H=\partial_{\phi}H=0$. From Equation (\ref{Ham}), the Hamilton--Jacobi equation can be written as
\begin{eqnarray}
0&=&-\frac{1}{\Delta}\left[a\frac{\partial S}{\partial \phi}+(r^2+a^2)\frac{\partial S}{\partial t}\right]^2 \nonumber\\
&+&\left[\frac{1}{\sin\theta}\frac{\partial S}{\partial \phi}+a\sin\theta\frac{\partial S}{\partial t}\right]^2 \\
&+&\left(\frac{\partial S}{\partial \theta}\right)^2+\Delta\left(\frac{\partial S}{\partial r}\right)^2+\rho^2\omega_p^2,~\label{HJ1}\nonumber
\end{eqnarray}
where $S$ is the Jacobi action with the following separable ansatz:
\begin{eqnarray}
S=-Et+L\phi+S_r(r)+S_{\theta}(\theta).~\label{Spaction}
\end{eqnarray}

Inserting Equation~(\ref{Spaction}) into Equation~(\ref{HJ1}), one can obtain
\begin{eqnarray}
0\!\!\!&=&\!\!\!-\frac{1}{\Delta}\left(a L+(r^2+a^2)E\right)^2+\left(\frac{L}{\sin\theta}-aE\sin\theta\right)^2 \nonumber\\
\!\!\!&+&\!\!\!\left(\partial_{\theta}S_{\theta}\right)^2+\Delta\left(\partial_rS_r\right)^2+\omega_p^2(r^2+a^2\cos^2\theta) \\
\!\!\!&=&\!\!\!-\frac{1}{\Delta}\left(a L+(r^2+a^2)E\right)^2+\left(\frac{L^2}{\sin^2\theta}-a^2E^2\right)\cos^2\theta \nonumber\\
\!\!\!&+&\!\!\!(L\!-\!aE)^2\!\!+\!\left(\partial_{\theta}S_{\theta}\right)^2\!\!+\!\Delta\!\!\left(\partial_rS_r\right)^2 \!+\!\omega_p^2(r^2\!+\!a^2\cos^2\theta). ~~~\nonumber
~\label{HJ2}
\end{eqnarray}

It is obvious that the Hamilton--Jacobi equation can be separated only provided that the plasma frequency takes the following form~\cite{Perlick2017}:
\begin{eqnarray}
\omega_p(r,\theta)^2=\frac{f_r(r)+f_{\theta}(\theta)}{r^2+a^2\cos^2\theta},
\end{eqnarray}
where $f_r(r)$ and $f_{\theta}(\theta)$ are some functions of $r$ and $\theta$, respectively. Then, Equation~(\ref{HJ2}) can be rewritten as
\begin{eqnarray}
&&\Delta\left(\partial_rS_r\right)^2-\frac{1}{\Delta}\left(a L+(r^2+a^2)E\right)^2+(L-aE)^2+f_r \nonumber\\
&&=-\left(\partial_{\theta}S_{\theta}\right)^2-\left(\frac{L^2}{\sin^2\theta}-a^2E^2\right)\cos^2\theta-f_{\theta}=\kappa.
\end{eqnarray}

It is clear that the first expression is independent of $\theta$, the second one is independent of $r$, and $\kappa$ is independent both of $r$ and $\theta$, thus $\kappa$ is a constant, which is related to the Carter constant.
Then, with the relations $p_r=\partial_rS_r$, $p_{\theta}=\partial_{\theta}S_{\theta}$, and $\dot{x}^{\mu}=\frac{\partial H}{\partial p_{\mu}}$, the equations of motion for the photon in the presence of the plasma are~\cite{Perlick2017,Yan2019}
\begin{eqnarray}
\rho^2 \frac{\partial t}{\partial \lambda}\!\!\!&=&\!\!\!\frac{1}{\Delta}\left[\big(2a^2M r \sin^2\theta\!+\!(a^2\!+\!r^2)\rho^2\big)E\!-\!2aMrL\right], ~~~~~\\
\rho^2 \frac{\partial \phi}{\partial \lambda}\!\!\!&=&\!\!\!\frac{2aMrE\sin^2\theta}{\Delta}+\frac{L(\rho^2-2Mr)}{\Delta\sin^2\theta}, \\
\rho^2 \frac{\partial r}{\partial \lambda}\!\!\!&=&\!\!\!\pm\sqrt{\mathcal{R}_p}, \\
\rho^2 \frac{\partial \theta}{\partial \lambda}\!\!\!&=&\!\!\!\pm\sqrt{\Theta_p},
\end{eqnarray}
where
\begin{eqnarray}
\mathcal{R}_p\!\!\!&=&\!\!\!\left[a L\!-\!E \left(a^2\!+\!r^2\right)\right]^2\!-\!\Delta\left[(L\!-\!a E)^2\!+\!\kappa\right]\!-\!f_r\Delta, ~~~ \\
\Theta_p\!\!\!&=&\!\!\!\kappa+a^2 E^2 \cos^2\theta-L^2 \cot^2\theta-f_{\theta}.
\end{eqnarray}

Following the same procedure as in Section~\ref{sectionB}, the unstable circular photon orbits should satisfy
\begin{eqnarray}
\mathcal{R}_p(r)=\frac{d\mathcal{R}_p(r)}{dr}=0.
\end{eqnarray}

Then, {we can obtain}
\begin{eqnarray}
\xi_p\!\!\equiv\!\!\frac{L}{\omega_0}\!\!\!&=&\!\!\!\frac{a \left(a^2\!+\!r^2\right) \Delta '\!-\!2 a r \Delta\!-\!\sqrt{a^2 \Delta^2 \left(4 r^2\!-\!\hat{f}_r' \Delta '\right)}}{a^2 \Delta '}, ~~~~~\\
\eta_p\!\!\equiv\!\!\frac{\mathcal{L}}{\omega_0^2}\!\!\!&=&\!\!\!-\frac{a^2 \hat{f}_r+r^4}{a^2}+\frac{\Delta\hat{f}_r'  \left(\Delta -a^2\right)}{a^2 \Delta '} \nonumber\\
\!\!\!&+&\!\!\!\frac{\Delta  \left(4 a^2 r\!+\!2 r^2 \Delta '\!-\!4 \Delta  r\right) \left(\sqrt{4 r^2\!-\!\Delta ' \hat{f}_r'}\!+\!2 r\right)}{a^2 \Delta '^2},
\end{eqnarray}
where $\hat{f}_r\equiv \frac{f_r(r)}{\omega_0^2}$ and we have introduced the photon energy $E=\omega_0$ with $\hbar=1$ for convenience.
With the presence of the plasma, the celestial coordinates are modified as~\cite{Perlick2017}
\begin{eqnarray}
x_p&=&-\xi_p\csc\theta_o,\\
y_p&=&\pm\sqrt{\eta_p+a^2\cos^2\theta_o-\xi_p^2\cot^2\theta_o-\hat{f}_{\theta_o}}.
\end{eqnarray}
where $\hat{f}_{\theta}\equiv \frac{f_{\theta}(\theta)}{\omega_0^2}$ and $\theta_o$ is the angular coordinate of the observer. To investigate the effects of the plasma on the shadow of the black hole, we choose $f_r(r)=\omega_c^2Mr$ and $f_{\theta}(\theta)=0$ as an example, i.e., the plasma frequency has the following form:
\begin{eqnarray}
\omega_p^2=\frac{\omega_c^2Mr}{r^2+a^2\sin^2\theta},
\end{eqnarray}
where $\omega_c$ is a constant. For later convenience, defining $k\equiv \frac{\omega_c}{\omega_0}$, we have $\hat{f}_r=k^2 M r$ and $\hat{f}_{\theta}=0$.

\begin{figure*}[htb]
\begin{center}
\subfigure[$~$]  {\label{sap}
\includegraphics[width=4.4cm]{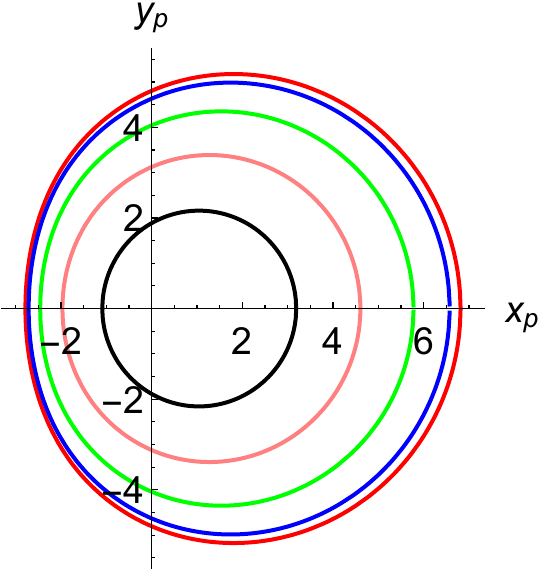}}
\subfigure[$~$]  {\label{dsap}
\includegraphics[width=4.4cm]{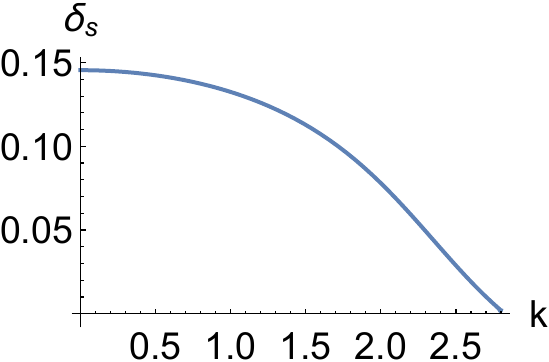}}
\subfigure[$~$]  {\label{Rska}
\includegraphics[width=4.4cm]{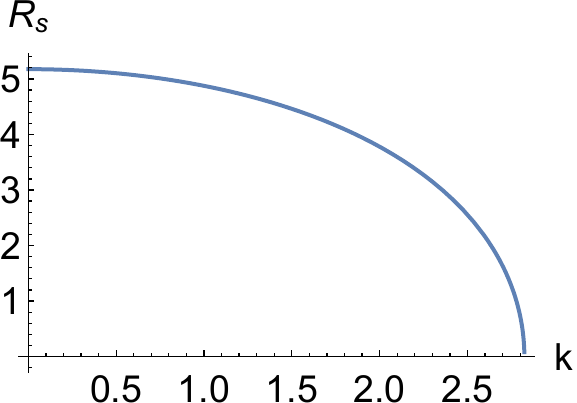}}
\end{center}
\caption{{The case for a rotating black hole surrounded by a plasma in model A. Panels (\textbf{a}), (\textbf{b}), and (\textbf{c}) plot the shadow, the distortion parameter, and the radius of the black hole, respectively.}  Here, $\theta_o=\frac{\pi}{2}$, $a=0.9$, $l=0.01$, and $k=0$ for red circle, $k=0.8$ for blue circle, $k=1.6$ for green circle, $k=2.2$ for pink circle, $k=2.6$ for black circle. When $k=k_{\text{max}}=2.823$, the shadow shrinks to a point.}

\label{spa}
\end{figure*}

\begin{figure*}[htb]
\begin{center}
\subfigure[$~$]  {\label{shbp}
\includegraphics[width=4.4cm]{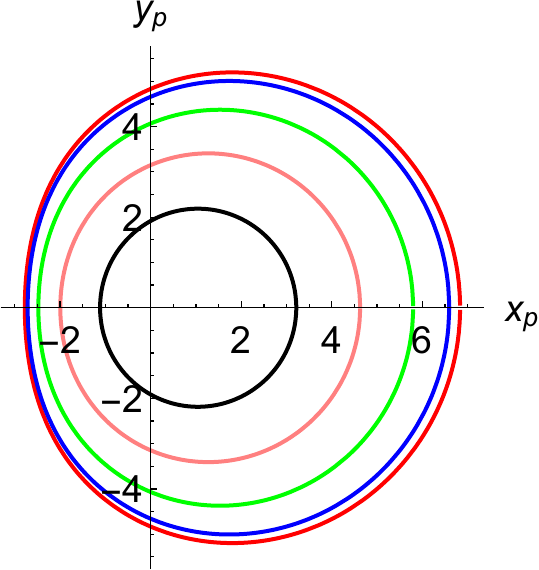}}
\subfigure[$~$]  {\label{dsbp}
\includegraphics[width=4.4cm]{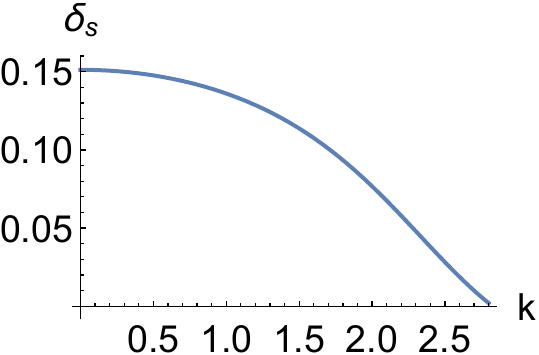}}
\subfigure[$~$]  {\label{Rskb}
\includegraphics[width=4.4cm]{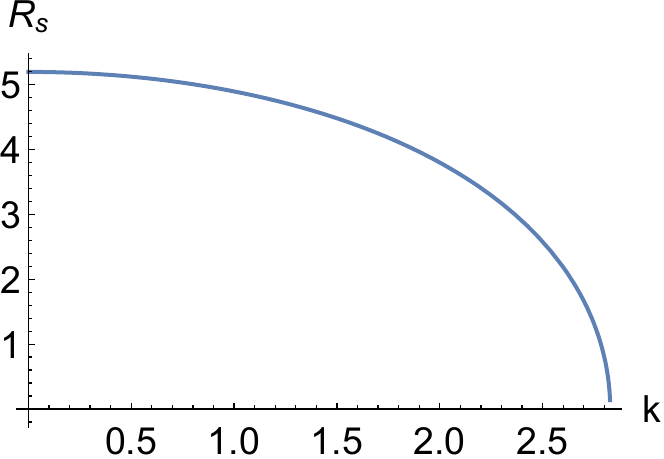}}
\end{center}
\caption{{The case for a rotating black hole surrounded by a plasma in model B. Panels (\textbf{a}), (\textbf{b}), and (\textbf{c}) plot the shadow, the distortion parameter, and the radius of the black hole, respectively.}  Here, $\theta_o=\frac{\pi}{2}$, $a=0.6$, $\beta=0.05$, and $k=0$ for red circle, $k=0.8$ for blue circle, $k=1.6$ for green circle, $k=2.2$ for pink circle, $k=2.6$ for black circle. When $k=k_{\text{max}}=2.828$, the shadow shrinks to a point.}
\label{spb}
\end{figure*}

Figures \ref{spa} and \ref{spb} plot the shadow of the black hole surrounded by the plasma for model A and model B, respectively. It is clear that the size of the silhouette decreases with $k$ for both models. Additionally, Figures \ref{spa}b and \ref{spb}b show that the distortion parameter $\delta_s$ decreases with $k$ and vanishes when $k$ approaches its maximum $k_{\text{max}}$, which indicates that the deformed shadow recovers to the standard circle at $k=k_{\text{max}}$. \mbox{Figures \ref{spa}c and \ref{spb}c} show that the shadow shrinks to a point when $k=k_{\text{max}}$. Since at this maximum the photon region for light rays crossing the equatorial plane vanishes, observers close to the equatorial plane no longer see a shadow.


{
\section{Gravitational Deflection of Light by Black Hole}

In this section, we will give a brief analysis about the weak deflection angle of light by rotating nonlocally modified black holes given in the above sections with the Gauss--Bonnet theorem~\cite{Gibbons235009,Werner44}. It is well known that the rotating black hole spacetime (\ref{rsol}) gives rise to a Finslerian optical metric of Randers type~\cite{Werner44}, with a positive definite Hessian
\begin{eqnarray}
g_{ij}=\frac{1}{2}\frac{\partial^2\mathcal{F}^2(x,v)}{\partial v^i \partial v^j}, ~\label{mofR}
\end{eqnarray}
and the Randers metric can be written as follows:
\begin{eqnarray}
\mathcal{F}(x,v)=\sqrt{\alpha_{ij}v^i v^j}+\beta_i(x)v^i,
\end{eqnarray}
where $\alpha_{ij}$ denotes the Riemannian metric and $\beta_i$ is a one-form satisfying $\alpha_{ij}\beta_i\beta_j<1$. After some simple calculations, one can obtain the following expression for the Randers metric:
\begin{eqnarray}
\mathcal{F}\left(r,\phi,\frac{dr}{dt},\frac{d\phi}{dt}\right)&=&\sqrt{\frac{r^4\left(\frac{dr}{dt}\right)^2}{\Delta(\Delta-a^2)}+\frac{r^4\Delta\left(\frac{d\phi}{dt}\right)^2}{(\Delta-a^2)^2}} \nonumber\\
&-&\frac{2M ar \mathcal{G}}{\Delta-a^2}\left(\frac{d\phi}{dt}\right),~~~\label{mofF}
\end{eqnarray}
where we have set $\theta=\frac{\pi}{2}$ without loss of generality. Then, one can construct a Riemannian manifold ($\mathcal{M},\bar{g}$) osculating the Randers manifold ($\mathcal{M},\mathcal{F}$) by applying Naz$\text{\i}$m's method. This can be carried out by choosing a smooth and nonzero vector field $\bar{v}$ over $\mathcal{M}$ with the definition $\bar{g}_{ij}(x)=g_{ij}(x,\bar{v}(x))$. Since we are only interested in the leading terms of the weak deflection angle, it suffices to take $r(\phi)=b/\sin\phi$ as the zero-order approximation of the deflected light and use the leading terms of the vector field $\bar{v}^r=-\cos\phi$ and $\bar{v}^{\phi}=\sin^2\phi/b$~\cite{Werner44}, where $b$ is the impact parameter.

Considering a nonsingular and simply connected domain $D_R$, which is bounded by the light ray $\gamma_{\bar{g}}$ and a circular curve $\gamma_R$ of radius $R$ centered on the lens, in the equatorial plane of the osculating Riemannian manifold defined above, the Gauss--Bonnet theorem states that~\cite{Gibbons235009,Werner44}
\begin{eqnarray}
\int\int_{D_R}K dS+\oint_{\partial D_R}\kappa dt+\sum_i \theta_i=2\pi \chi(D_R), ~\label{GBT}
\end{eqnarray}
where $K$ is the Gaussian curvature of the domain $D_R$, $\kappa$ is the geodesic curvature of the boundary defined as $\kappa=|\nabla_{\dot{\gamma}}\dot{\gamma}|$, and $\theta_i$ stands for the $i^{\text{th}}$ exterior angles. For the light ray $\gamma_{\bar{g}}$ intersecting the circular curve $\gamma_R$ in the observer $O$ and the source $S$, there are only two exterior angles, $\theta_{O}$ and $\theta_S$. Letting $R\rightarrow\infty$, the exterior angles yield $\theta_{O}+\theta_{S}\rightarrow\pi$. In addition, in this limit, the geodesic curvature of the circular curve reduces to $\kappa(\gamma_R)\rightarrow R^{-1}$ and $dt\rightarrow R d\phi$. Then, Equation~(\ref{GBT}) becomes
\begin{eqnarray}
\int\int_{D_{\infty}}K dS+\int_0^{\pi+\alpha}d\phi=\pi,
\end{eqnarray}
where we have used $\kappa(\gamma_{\bar{g}})=0$ and the Euler characteristic $\chi(D_R)=1$. Then, the weak deflection angle $\alpha$ can be formally expressed as
\begin{eqnarray}
\alpha=-\int\int_{D_{\infty}}K dS.
\end{eqnarray}

Let us now compute the metric of the osculating Riemannian manifold. From  \mbox{Equations~(\ref{mofR}) and (\ref{mofF})}, we have
\begin{eqnarray}
\bar{g}_{rr}&=&1+\frac{4 M \mathcal{G}}{r}-\frac{2 a M r \mathcal{G} \sin^6\phi}{b^3 \left(\cos ^2\phi+\frac{r^2}{b^2}\sin ^4\phi\right)^{3/2}} \nonumber\\
&+&\mathcal{O}(M^2,a^2), \\
\bar{g}_{r\phi}&=&\frac{2 a M \mathcal{G} \cos ^3\phi }{r \left(\cos ^2\phi+\frac{r^2 }{b^2}\sin ^4\phi \right)^{3/2}}+\mathcal{O}(M^2,a^2),\\
\bar{g}_{\phi\phi}&=&r^2+2 M r \mathcal{G}-\frac{2 a M r \mathcal{G} \sin ^2\phi}{b \left(\cos ^2\phi+\frac{r^2 }{b^2}\sin ^4\phi \right)^{3/2}}\bigg(3 \cos ^2\phi \nonumber\\
&+&\frac{2 r^2 }{b^2}\sin ^4\phi\bigg)+\mathcal{O}(M^2,a^2),
\end{eqnarray}
with the determinant given as
\begin{eqnarray}~\label{detg}
\det\bar{g}=r^2+6 M r \mathcal{G}-\frac{6 a M r \mathcal{G} \sin ^2\phi }{b \sqrt{\cos ^2\phi +\frac{r^2 }{b^2}\sin ^4\phi}}.
\end{eqnarray}

The corresponding Gaussian curvature of the domain $D_{\infty}$ is
\begin{eqnarray}
K&=&\frac{\bar{R}_{r\phi r\phi}}{\det \bar{g}}=-\frac{M \left(r \left(r \mathcal{G}''-2 \mathcal{G}'\right)+2 \mathcal{G}\right)}{r^3}+f(r,\phi) \nonumber\\
&+&\mathcal{O}(M^2,a^2), ~\label{GC}
\end{eqnarray}
where
\begin{widetext}
\begin{eqnarray}
f(r,\phi)&=&\frac{3 a M \mathcal{G} \sin ^3\phi}{r^2 b^7 \left(\cos ^2\phi+\frac{r^2\phi}{b^2}\sin ^4\right)^{7/2}}\bigg[2 b^5 \cos ^6\phi \left(-2+\frac{5 r}{b}\sin \phi\right)+b^3 r^2 \bigg(\sin ^4( 2 \phi ) \nonumber\\
&-&\frac{r}{b}\sin ^9\phi+\frac{2 r^3 }{b^3}\sin ^{11}\phi\bigg)+b^5 \sin ^2\phi  \cos ^4\phi  \left(-2+\frac{9 r }{b}\sin \phi-\frac{8 r^3 }{b^3}\sin ^3\phi\right) \nonumber\\
&+&b^4 r \sin ^5\phi  \cos ^2\phi  \bigg(4+\frac{8 r }{b}\sin \phi-\frac{3 r^2}{b^2}+\frac{r^2}{b^2}\cos (2 \phi )\bigg)\bigg] \\
&+&\frac{a M \sin ^2\phi }{b^2 r \left(\cos ^2\phi+\frac{r^2 }{b^2}\sin ^4\phi \right)^{5/2}}\bigg[b \mathcal{G}'' \bigg(3 \cos ^4\phi+\frac{5 r^2 }{b^2}\sin ^4\phi  \cos ^2\phi+\frac{2 r^4 }{b^4}\sin ^8\phi \bigg) \nonumber\\
&-&\mathcal{G}'\sin \phi  \bigg(12 \cos ^4\phi+\sin ^2\phi  \cos ^2\phi  \left(6+\frac{11 r }{b}\sin \phi\right)+\frac{5 r^3 }{b^3}\sin ^7\phi \bigg)\bigg].\nonumber
\end{eqnarray}
\end{widetext}

Then, the weak deflection angle of light by the rotating nonlocally modified black holes obtained in the above sections can be calculated as
\begin{eqnarray}
\alpha_A\!\!&=&\!\!\frac{4M}{b}\pm\frac{4aM}{b^2}+\frac{6M l}{b^2}+\mathcal{O}(M^2,a^2,l^2,Mal), \\
\alpha_B\!\!&=&\!\!\frac{4M}{b}\pm\frac{4aM}{b^2}-\frac{16M\beta}{3b^3}+\mathcal{O}(M^2,a^2,\beta^2,Ma\beta),~~~
\end{eqnarray}
where the positive and negative signs represent the retrograde and prograde light rays, respectively. It is obvious that the {nonlocal} parameter $l$ makes a positive contribution to the deflection angle and its contribution is larger than the rotation parameter $a$. The {nonlocal} parameter $\beta$ makes a negative contribution and is suppressed by $b^{-3}$, thus its influence can be ignored compared with the rotation parameter $a$.
}

\section{Conclusions}~\label{con}
In this paper, we constructed two kinds of rotating black hole solutions in the nonlocally modified gravitational theory by using the Newman--Janis algorithm without complexification. {We first investigated the effects of the nonlocal correction on the black hole shadow by analyzing the null geodesics of the spacetime of these two types of rotating black holes, and then calculated the weak deflection angle of light by these two rotating black holes with the Gauss--Bonnet theorem.}

Without loss of generality, we assume the observer is located at the equatorial plane of the black hole. It was found that the size of the shadow decreases with the {nonlocal} parameters. {The reason is that the strength of the gravitational interaction decreases with the nonlocal parameters at a fixed proper distance to the center of the black hole, which results in a smaller escape velocity. A smaller escape velocity will lead to a more compact event horizon.} Furthermore, {since the prograde photon experiences a smaller unstable circular orbit compared with the retrograde photon, the black hole shadow is no longer circular. Since the radius of the smaller unstable circular orbit decreases with the rotation parameter while the radius of the larger unstable circular orbit increases with the rotation parameter, the black hole shadow shifts to the right with the increasing rotation parameter. As the radius of the prograde orbit decreases faster than the retrograde orbit increases, the shape of the shadow becomes more deformed with the rotation parameter. However, the size of the black hole shadow almost does not change with the rotation parameter since the rotation parameter has no influence on the strength of the gravitational force.} In addition, the shapes of the shadow remain approximately circular when the rotation parameter is small even if the {nonlocal} parameter approaches its extremum.

Then, we studied the radius and distortion of the black hole shadow in both models. The results showed that the radius of both models decreases with the nonlocal parameters because of the decrease in the strength of the gravitational interaction, which is in consistence with the former analysis. Additionally, the distortions for both models increase with the nonlocal parameters and the shadow becomes more deformed with larger rotation parameter for fixed nonlocal parameters. Assuming the area of the black hole shadow is equal to the high-energy absorption cross section, the energy emission rate was also investigated. It was found that there exists a peak for each curve and the peak decreases and shifts to the low frequency with the increasing nonlocal parameters. In addition, the peak also decreases with the rotation parameter for fixed nonlocal parameters, since the area of the event horizon decreases with the nonlocal parameters.

Next, we also discussed the effects of the plasma on the black hole shadow. It was shown that the size of the shadow decreases with the presence of plasma for both models. Additionally, the deformed shadow recovers to the standard circle gradually with increasing $k$ and shrinks to a point at the maximum $k_{\text{max}}$, which is because the photon region for light rays crossing the equatorial plane vanishes, i.e., the shadow becomes invisible for the observer close to the equatorial plane.

Finally, we calculated the weak deflection angle of light by these two rotating black holes and analyzed the effects of the nonlocal correction on the deflection angle. It was found that the nonlocal parameter of model A makes a positive contribution to the deflection angle, which can be compared with that of the angular momentum of the black hole. The nonlocal parameter of model B makes a negative contribution. However, its influence is suppressed by $b^{-3}$ and thus can be ignored compared with the rotation parameter.

{In this paper, the nonlocality led by the Lagrangian $\mathcal{A}^{-2}(\square)R$ is discussed. Actually, the nonlocality can also be led by a quadratic term such as $Rf(\square)R$ in the Lagrangian. This is also an important case and deserves to be discussed in detail in the same framework. We will make a detailed analysis for the case of nonlocality from quadratic term in a \mbox{future work}.}

\acknowledgments{
This work was supported by the National Natural Science Foundation of China (Grants Nos. 11875151, 12075103, 11975072, 11835009, 11875102, and 11690021), the Liaoning Revitalization Talents Program (Grant No. XLYC1905011), the Fundamental Research Funds for the Central Universities (Grant No. N2005030), the National Program for Support of Top-Notch Young Professionals (Grant No. W02070050), the Science Research Grants from the China Manned Space Project (Grant No. CMS-CSST-2021-B01), the National 111 Project of China (Grant No. B16009), and the Education Department of Shaanxi Province (No. 20JK0553).}

\end{document}